\documentclass[twocolumn]{aastex63}
\usepackage[utf8]{inputenc}
\usepackage{natbib}
\setcitestyle{round}
\usepackage{amsmath,amssymb}
\usepackage{afterpage}
\usepackage{xcolor}
\usepackage[caption=false]{subfig}
\usepackage{soul}
\usepackage{enumitem}
\usepackage{xcolor}

\newcommand{\hi}{{\rm H\,{\small I}}}
\newcommand{\hii}{{\rm H\,{\small II}}}
\newcommand{\kms}{\ensuremath{\,{\rm km\,s^{-1}}}}

\newcommand{\percc}{\ensuremath{\,{\rm cm^{-3}}}}
\newcommand{\ts}{\ensuremath{T_s}}
\newcommand{\nhi}{\ensuremath{N(\hi{})}}
\newcommand{\persc}{\ensuremath{\,{\rm cm^{-2}}}}
\newcommand{\hcop}{\text{HCO\textsuperscript{+}}}
\newcommand{\hcn}{\text{HCN}}
\newcommand{\hnc}{\text{HNC}}
\newcommand{\cch}{\text{C\textsubscript{2}H}}
\newcommand{\htwo}{\text{H\textsubscript{2}}}

\defcitealias{2021arXiv210906273R}{Paper I}

\shorttitle{Atomic conditions suitable for molecule formation: Comparison to models}
\shortauthors{Rybarczyk et al.}

\begin{document}

\title{The role of neutral hydrogen in setting the abundances of molecular species in the Milky Way's diffuse interstellar medium. II. Comparison between observations and theoretical models}

\correspondingauthor{Daniel R. Rybarczyk}
\email{rybarczyk@astro.wisc.edu}

\author[0000-0003-3351-6831]{Daniel R. Rybarczyk}
\affiliation{University of Wisconsin--Madison, Department of Astronomy, 475 N Charter St, Madison, WI 53703, USA}

\author[0000-0003-1613-6263]{Munan Gong}
\affiliation{Max-Planck-Institut f\"ur Extraterrestrische Physik, Garching by Munich, D-85748, Germany}

\author[0000-0002-3418-7817]{Sne\v zana Stanimirovi\'c}
\affiliation{University of Wisconsin--Madison, Department of Astronomy, 475 N Charter St, Madison, WI 53703, USA}

\author[0000-0002-6984-5752]{Brian Babler}
\affiliation{University of Wisconsin--Madison, Department of Astronomy, 475 N Charter St, Madison, WI 53703, USA}

\author[0000-0002-7743-8129]{Claire E. Murray}
\affiliation{Department of Physics \& Astronomy, Johns Hopkins University, 3400 N. Charles Street, Baltimore, MD 21218, USA}

\author[0000-0001-6114-9173]{Jan Martin Winters}
\affiliation{Institut de Radioastronomie Millim\'etrique (IRAM), 300 rue de la Piscine,  F-38406 St. Martin d’H\'eres, France}

\author[0000-0002-1583-8514]{Gan Luo}
\affiliation{School of Astronomy and Space Science, Nanjing University, Nanjing 210093, People’s Republic of China}
\affiliation{Key Laboratory of Modern Astronomy and Astrophysics (Nanjing University), Ministry of Education, Nanjing 210093, People’s Republic of China}

\author[0000-0003-0109-2392]{T. M. Dame}
\affiliation{Center for Astrophysics | Harvard \& Smithsonian, 60 Garden St
Cambridge, MA 02138, USA}

\author[0000-0002-9512-5492]{Lucille Steffes}
\affiliation{University of Wisconsin--Madison, Department of Astronomy, 475 N Charter St, Madison, WI 53703, USA}
\begin{abstract}

We compare observations of \hi{} from the Very Large Array (VLA) and the Arecibo Observatory and observations of \hcop{} from the Atacama Large Millimeter/submillimeter Array (ALMA) and the Northern Extended Millimeter Array (NOEMA) in the diffuse ($A_V\lesssim1$) interstellar medium (ISM) to predictions from a photodissociation region (PDR) chemical model and multi-phase ISM simulations. Using a coarse grid of PDR models, we estimate the density, FUV radiation field, and cosmic ray ionization rate (CRIR) for each structure identified in \hcop{} and \hi{} absorption. These structures fall into two categories. Structures with $T_s<40$ K, mostly with $N(\hcop{})\lesssim10^{12}$ \persc{}, are consistent with modest density, FUV radiation field, and CRIR models, typical of the diffuse molecular ISM. Structures with spin temperature $T_s>40$ K, mostly with $N(\hcop{})\gtrsim10^{12}$ \persc{}, are consistent with high density, FUV radiation field, and CRIR models, characteristic of environments close to massive star formation. The latter are also found in directions with a significant fraction of thermally unstable \hi{}. In at least one case, we rule out the PDR model parameters, suggesting that alternative mechanisms (e.g., non-equilibrium processes like turbulent dissipation and/or shocks) are required to explain the observed \hcop{} in this direction. Similarly, while our observations and simulations of the turbulent, multi-phase ISM agree that \hcop{} formation occurs along sightlines with $N(\hi{})\gtrsim10^{21}$ \persc{}, the simulated data fail to explain \hcop{} column densities $\gtrsim\rm{few}\times10^{12}$ \persc{}. Since a majority of our sightlines with \hcop{} had such high column densities, this likely indicates that non-equilibrium chemistry is important for these lines of sight.

\end{abstract}

\section{Introduction} \label{sec:intro}

Many molecular species have been detected in the diffuse interstellar medium (ISM), revealing important chemistry even at $A_V\lesssim1$  \citep[and references therein]{1991ApJ...371L..77M,1996A&A...307..237L,2000A&A...355..327L,2000A&A...358.1069L,2001A&A...370..576L,2014A&A...564A..64L,2006ARA&A..44..367S}. \hcop{} is one of the most commonly detected molecules in the diffuse ISM \citep[e.g.,][]{1996A&A...307..237L} and has been shown to be an excellent tracer of molecular hydrogen, \htwo{} \citep{2000A&A...355..333L,2010A&A...518A..45L}. However, an outstanding problem for the last few decades has been that the observed  \hcop{} column densities in the diffuse ISM are often one to two orders of magnitude higher than what is expected theoretically from UV-dominated chemical models \citep{1996A&A...307..237L}, with large column density variations observed across interstellar clouds of similar total hydrogen column density \citep[see, e.g., Figure 3 of][hereafter \citetalias{2021arXiv210906273R}]{2021arXiv210906273R}.

\hcop{} has been observed in absorption in the direction of extragalactic point sources \citep[e.g.,][]{1996A&A...307..237L,2018A&A...610A..49L,2010A&A...520A..20G,2018A&A...617A..54L,2019A&A...622A..26G,2020ApJ...889L...4L} as well as Galactic \hii{} regions \citep[e.g.,][]{1983A&A...120..307N,2001A&A...371..287M,2010A&A...513A...9C,2019A&A...622A..26G}. The range of the observed \hcop{} column  densities spans $\sim\text{few}\times10^{10}$ \persc{} to $\sim10^{14}$ \persc{}, with an essentially continuous distribution sampling a range of interstellar environments, from more diffuse molecular gas to dense dark clouds. More diffuse environments ($A_V\lesssim1$) typically have \hcop{} column densities $<10^{12}$ \persc{}, and often show weak CO emission \citep{1998A&A...339..561L}, while clouds in the Galactic Center or prominent local clouds such as Taurus or California have column densities $>10^{12}$ \persc{} and tend to have more pronounced CO emission.  

Several previous studies have modeled \hcop{} observations using photodissociation region (PDR) chemical models, where FUV photons are attenuated by dust and molecules as $A_V$ increases, allowing molecular species to form in shielded regions, and where molecular abundances are calculated under the steady state assumption \citep[e.g.,][]{HT1999}. Such models have often been able to explain the column densities observed in denser environments, but have routinely failed to explain the high \hcop{} column densities observed in the diffuse ISM \citep[e.g.,][]{2009A&A...495..847G}. 
This suggests that some key elements, likely involving dynamical effects, are still not accounted for in our conventional understanding of the formation and evolution of molecules in the diffuse ISM \citep{2005IAUS..231..187L,2009A&A...495..847G,Valdivia2017,Lesaffre2020}.

Several alternative chemical models have been proposed to explain the enhanced column densities of species such as \hcop{} in the diffuse ISM. For example, the turbulent mixing between the cold neutral medium (CNM) and warm neutral medium (WNM) \citep{Lesaffre2007}, or turbulent dissipation in shocks or velocity shear \citep{2003MNRAS.343..390F,2006A&A...452..511F,2009A&A...495..847G,2020A&A...643A.101L}. In particular, the turbulent dissipation region (TDR) model has had success is explaining high \hcop{} column densities in the diffuse ISM \citep{2009A&A...495..847G}. In this model, the release of suprathermal energy by turbulent dissipation enhances the formation of various molecular species along the line of sight. The model predictions were strongly dependent on the gas density and the turbulent rate of strain, but for a wide range of values, the TDR model was able to explain the high \hcop{} and CH$^+$ column densities observed in the diffuse ISM. Although several species were still underpredicted by an order of magnitude and the results are also dependent on the chemistry, this nevertheless demonstrates that attention to dynamical processes can significantly reduce the disparity between model predictions and observational results. More recently, intermittent turbulent dissipation---which occurs on timescales much shorter than the $10^6$--$10^7$ yr required for \htwo{} to reach steady state in the diffuse ISM \citep{GoldsmithLi2005}---has been proposed to explain the CH$^+$ column densities in the diffuse ISM that far exceed PDR model predictions  \citep[e.g.,][]{Valdivia2017,Lesaffre2020}. As CH$^+$ contributes to \hcop{} formation, this may suggest that the high \hcop{} column densities observed in the diffuse ISM reflect a large supply of CH$^+$ from dynamical events. 

These alternative chemical models couple the chemical evolution of gas with the turbulent dynamical evolution of the environment. Therefore, to distinguish between these different models we need to trace the underlying dynamical processes which influence molecule formation. With the goal of diagnosing the underlying properties of atomic gas, which is the main ingredient for the formation and survival of molecules, we recently conducted a survey of \hcn{}, \cch{}, \hcop{}, and \hnc{} in the direction of 20 background radio continuum sources where the 21-SPONGE project \citep[21 cm Spectral Line Observations of Neutral Gas with the Karl G. Jansky Very Large Array;][]{2015ApJ...804...89M,2018ApJS..238...14M} previously observed atomic hydrogen (\hi{}) in emission and absorption \citepalias{2021arXiv210906273R}.With the characterization of the line-of-sight environments from 21-SPONGE, we can highlight the interstellar conditions where models typically fail to match observed molecular column densities. This approach has not been used previously, yet it can differentiate the environments where UV-dominated chemical models are sufficient to explain the observed molecular column densities from those where alternative chemical models are needed. Specifically, in this work we test whether the \citet{Gong2017} PDR model can reproduce the observed \hcop{} column densities given the environmental constraints from 21-SPONGE for absorption features detected in \citetalias{2021arXiv210906273R}. We use a set of models with a broader range of far-ultraviolet (FUV) interstellar radiation field (ISRF) strengths and cosmic ray ionization rates (CRIR) than have previously been employed in comparisons of PDR model predictions to \hcop{} abundances in the diffuse ISM \citep{2010A&A...520A..20G}. We do not seek a precise model fit to the data, but rather a rough estimate of the density, FUV radiation field, and CRIR needed by the PDR model to explain observations. We then compare these estimates to existing observational probes of these environmental parameters along the observed lines of sight. Sightlines where the PDR estimates differ from observationally constrained estimates by an order of magnitude or more likely mark sites of non-equilibrium chemistry, where the PDR models fail to explain \hcop{} abundances due to the role of dynamical processes in molecule formation and survival.
Alternative chemical models are likely required to explain the \hcop{} column densities in such cases, although, as we discuss in Section \ref{sec:comparision_PDR} and Section \ref{section:comparison_tigress}, it may also be possible to understand high \hcop{} column densities for PDRs near \hii{} regions. We discuss further in Section \ref{sec:discussion} whether this explanation is plausible for our lines of sight.

Furthermore, despite the power of PDR models to explain observed molecular abundances in some (denser) interstellar environments, the PDR model remains a very crude representation of the ISM---the realistic ISM is turbulent and multi-phase, with complex dynamical, thermal, and chemical structures. This situation can be improved somewhat by pairing chemical modeling with magnetohydrodynamical (MHD) simulations, which provide a more realistic description of density structures in the turbulent, multi-phase ISM. For example, post-processing PDR chemistry on lines of sight from MHD simulations rather than uniform density cloud models can change the column density estimates by a factor of at least a few for key species in the diffuse ISM and bring theoretical predictions closer to observational results \citep{Levrier2012}. Such work often relies on simplified networks that reliably reproduce PDR model results at a lower computational cost than a full chemical network. More complex chemical modeling is also possible with MHD simulations \citep[e.g.,][]{Glover2010}, but this can be very computationally costly. A variety of different chemical models have emphasized that more realistic models of ISM density structures from MHD simulations significantly impact molecular abundances, and are  important for understanding the observed molecular column densities in the ISM \citep[e.g.,][]{Levrier2012,Gong2018}. Thus, in this work, we also compare the \hcop{} \citepalias{2021arXiv210906273R} and multiphase \hi{} \citep{2018ApJS..238...14M} column densities to predictions from chemically post-processed MHD simulations \citep{Gong2020}.

In \citetalias{2021arXiv210906273R}, we presented a summary of new observations of \cch{}, \hcn{}, \hcop{}, and \hnc{} in absorption obtained using the Atacama Large Millimeter/submillimeter Array (ALMA) and the Northern Extended Millimeter Array (NOEMA). We also discussed methods for extracting \hcop{} column densities and decomposing absorption spectra into Gaussian components. We used the complementary observations of atomic and molecular gas to investigate the atomic gas conditions necessary for molecule formation in the diffuse ISM. We demonstrated that \hcop{} (as well as \cch{}, \hcn{}, and \hnc{}) forms along sightlines where the visual extinction is $\gtrsim0.25$ and the column density of cold \hi{} is $\gtrsim10^{20}$ \persc{}, similar to the observed conditions required for the \hi{}-to-\htwo{} transition in the Galactic ISM \citep[e.g.,][]{1977ApJ...216..291S,Shull2021}. Moreover, we found that these molecular species were associated only with structures with an \hi{} optical depth $\gtrsim0.1$, spin temperature $<80$ K, and turbulent Mach number $\gtrsim2$. We note that the spin temperature is approximately equal to the kinetic temperature in these environments. Also, the higher turbulent Mach numbers, which are consistent with previous results \citep[e.g.,][]{Burkhart2015}, reflect the colder gas temperature rather than higher turbulent velocities in most cases. The turbulent Mach number is not used as a constraint in this work.

In this second paper, we compare the results from \citetalias{2021arXiv210906273R} with predictions from the \citet{Gong2017} PDR chemical model and the \citet{Gong2020} ISM simulations. In Section \ref{sec:observations}, we briefly introduce the data and methods presented in \citetalias{2021arXiv210906273R}, including observations of \hi{} emission and absorption from the 21-SPONGE project \citep{2015ApJ...804...89M,2018ApJS..238...14M}, new observations of \hcop{} absorption from ALMA and NOEMA, and supplementary archival measurements of dust extinction and dust temperature from the \textit{Planck} satellite \citep{2014A&A...571A..11P}. We then discuss the PDR model from \citet{Gong2017} and the ISM simulations from \citet{Gong2020} in Section \ref{sec:description_of_models}. In Section \ref{sec:comparision_PDR}, we find the PDR models that best reproduce the observed \hcop{} and \hi{} gas properties, from a grid of models with varying densities, FUV fields, and CRIRs. In Section \ref{section:comparison_tigress}, we compare the observed line of sight column densities to predictions derived from the multi-phase ISM simulations in \citet{Gong2020}. We then discuss our results in Section \ref{sec:discussion}. We compare the PDR model results in Section \ref{sec:comparision_PDR} (constraints on the local density, FUV field, and CRIR) to previous observations in these directions to test whether the PDR model results are plausible. Our conclusions are then presented in Section \ref{sec:conclusions}.

\section{Data} \label{sec:observations}

Here we provide a brief summary of new and existing observations used in this study. A full description of these observations was presented in \citetalias{2021arXiv210906273R}. 

\subsection{Observations of \hi{} with 21-SPONGE} \label{subsec:21-sponge}
The 21-SPONGE project \citep[][]{2015ApJ...804...89M,2018ApJS..238...14M} observed Galactic \hi{} absorption and emission with the Very Large Array (VLA) and the Arecibo Observatory in the direction of 57 bright background radio continuum sources with Galactic latitude $3.7^\circ\leq|b|\leq81^\circ$. 
The \hi{} absorption and emission spectra were decomposed into Gaussian components. The optical depth, $\tau$, spin temperature, \ts{}, and column density, \nhi{}, of each \hi{} component was estimated using the equations of radiative transfer. From the spin temperatures, \citet{2018ApJS..238...14M} also determined the fraction of \hi{} in the each of the three atomic gas phases in the neutral ISM: the cold neutral medium (CNM, $T_s<250~\rm{K}$), the unstable neutral medium (UNM, $250~\rm{K}<T_s<1000~\rm{K}$), and the warm neutral medium (WNM, $T_s<1000~\rm{K}$).
We use the \hi{} properties constrained by 21-SPONGE for comparison with the PDR model and MHD simulations.

\subsection{Observations of \hcop{} absorption with ALMA and NOEMA\label{subsec:hcop_obs}}
We observed \cch{}, \hcn{}, \hcop{}, and \hnc{} in absorption with ALMA during observing Cycles 6 and 7 (ALMA-SPONGE)
in the direction of 19 bright background radio continuum sources previously observed by the 21-SPONGE project. We observed an additional three sources (two overlapping with ALMA-SPONGE) in \hcop{} and \hcn{} absorption with NOEMA (NOEMA-SPONGE). Here we consider only the \hcop{} absorption spectra, as \cch{}, \hcn{}, and \hnc{} are not included in the \citet{Gong2017} chemical model.

We decomposed the \hcop{} absorption spectra into Gaussian components. The \hcop{} column density was calculated from the optical depth integral, $N(\hcop{})=1.11\times10^{12}~\persc{}~\int\tau(\hcop{})dv$, for individual components and for the total line of sight. This conversion assumes an excitation temperature equal to the temperature of the CMB, 2.725 K \citep{2010A&A...520A..20G,2020ApJ...889L...4L}.
We test and discuss this assumption in Section \ref{section:comparison_tigress}.

\subsubsection{Identifying the corresponding \hi{} spectral features}
In total, nine sightlines in \citetalias{2021arXiv210906273R} showed \hcop{} absorption at a level $\geq3\sigma$, comprising 23 Gaussian components. \citet{2018ApJS..238...14M} identified \hi{} absorption in all 20 sightlines, comprising 101 Gaussian components. Using a radial velocity matching criterion, we identified the \hi{} structures with a molecular component (i.e., those which were coincident in velocity with \hcop{} absorption). In some cases, a single \hi{} feature was associated with multiple \hcop{} features. In these cases, we consider the total \hcop{} column density (the sum of the individual \hcop{} column densities). We also ignore five features for which \citet{2018ApJS..238...14M} were unable to fit a physically meaningful $T_s$, since $T_s$ is an important constraint in our fitting.

We find 11 features with detailed \hi{} properties from 21-SPONGE and well-constrained \hcop{} column densities from \citetalias{2021arXiv210906273R}.
Table \ref{tab:PDR_model_fit} lists the observed parameters we use for PDR modeling for these structures---the \hcop{} column density, the \hi{} optical depth, and the \hi{} spin temperature (see Equation \ref{eq:chisq}). These parameters are also plotted in Figure \ref{fig:NHCOp_tauHI}. Figure \ref{fig:NHCOp_tauHI} also includes the remaining 85 components where \citet{2018ApJS..238...14M} detected \hi{} in absorption but we detected no \hcop{}; upper limits to the \hcop{} column densities for these features are shown in gray.

\subsection{$E(B-V)$ and $T_d$ from Planck} \label{subsec:planck}

We estimate the total hydrogen column density, $N_\mathrm{H}=N(\hi{})+2N(\htwo{})$, from the interstellar reddening, $E(B-V)$, 
\begin{equation} \label{eq:NH}
 N_\mathrm{H}=2.08\times10^{21}\times3.1 E(B-V) \text{ }\mathrm{cm^{-2}\,mag^{-1}},
\end{equation}
\citep{2017MNRAS.471.3494Z}. Here, $3.1E(B-V)$ is used as an estimate of the visual extinction, $A_V$. For each sightline, we extract $E(B-V)$, derived from the dust radiance measured by the \textit{Planck} satellite \citep{2014A&A...571A..11P}, from the nearest pixel using the dustmaps Python package \citep{2018JOSS....3..695M}. We also extract the dust temperature, $T_d$, from the maps derived by \citet{2016A&A...596A.109P} based on a modified blackbody spectral model of \textit{Planck} temperature maps at 353, 545, 857, and 3000 GHz. 
Since the dust temperature is set in part by the strength of the interstellar radiation field (ISRF), it can be used to estimate the strength of the radiation field, $\propto T_d^{\beta+4}$, where $\beta$ (also measured by \textit{Planck}) is the power law index that describes the dust emissivity cross-section as a function of frequency \citep[e.g., Equation 7.15 of][]{2005ism..book.....L}. However, the dust temperature also depends on the grain size distribution, grain composition, and structure, which results in a more complex relationship between $T_d$ and the strength of the ISRF. Moreover, the resolution of the \textit{Planck} dust maps is 5\arcmin{}, so the values of $E(B-V)$, $T_d$, $\beta$, and the ISRF strength should only be considered as very rough estimates to the local properties of the gas sampled by our \hi{} and molecular pencil-beam absorption spectra. We also note that the estimates of $T_d$, $\beta$, and the ISRF strength derived from \textit{Planck} observations are not included in our modeling; they are only discussed in Section \ref{sec:discussion} for context.

\section{Descriptions of Theoretical Models}
\label{sec:description_of_models}

We now briefly describe the PDR chemical model from \citet{Gong2017} and the ISM simulations from \citet{Gong2020}. In Sections \ref{sec:comparision_PDR} and \ref{section:comparison_tigress}, we compare the \hcop{} and \hi{} observations from \citetalias{2021arXiv210906273R} and \citet{2018ApJS..238...14M} to the predictions from these models.

\subsection{PDR Models}

We use the PDR code developed by \citet{Gong2017} which is publicly available \footnote{\url{https://sites.google.com/view/munangong/pdr-code}}. 
The simplified chemical network in the PDR code has been tested against a much larger network, and proved to produce accurate abundances of all modelled chemical species such as \hcop{}, CO, $\mathrm{H_2}$, $\mathrm{H_3^+}$, etc. \citep{Gong2017}.
We adapt a 1D slab model of uniform density, with a CRIR. 
The incident FUV radiation field comes from one side of the slab\footnote{We have also experimented with two-sided slab models, which we approximate by stitching two one-sided slabs back-to-back together, since the PDR code only support radiation coming from one side. We do not find any significant difference in our results.}, which is then attenuated by both dust and molecular shielding. 
The code uses a simplified chemical network that solves the evolution of chemical species such as \hcop{}, $\mathrm{CO}$, and $\mathrm{H_2}$. \hcn{}, \hnc{}, and \cch{}---which were also observed in \citetalias{2021arXiv210906273R}---are not yet included in the chemical network, so we focus only on the properties of \hcop{} and \hi{}. The heating and cooling of the gas are solved simultaneously with the chemical evolution. The solar neighborhood condition assumes a primary CRIR per H atom of $\xi=\xi_0$, where $\xi_0=2\times 10^{-16}~\mathrm{s^{-1}}$ is the measured local CRIR \citep{Indriolo2007}, and the FUV radiation field of $G'=1$, where $G'$ is the radiation field strength relative to the standard \citet{Draine1978} field. For each PDR model, we use a fixed value of $\xi$ and $G'$, run the chemistry, temperature, and radiation transfer up to a total column (in units of H atoms) of $N_\mathrm{H}=5\times 10^{22}~\mathrm{cm^{-2}}$,\footnote{This column density is set to be higher than the total column density derived from Planck dust emission along any of the observed sight-lines (see Section \ref{subsec:planck}).} and iterate until steady state. We vary the density of the PDR (in the unit of H atoms) between $n=10-10^4~\mathrm{cm^{-3}}$. To explore the effect of the environment, we vary the FUV radiation field $G'=0.1-100$ and $\xi/\xi_0=0.1-10$, with a very coarse grid, spacing the values
of $G'$ and $\xi$ by factors of 10. 

For each component of the Gaussian decomposition where \hcop{} is detected, there 
are three values constrained by the observations in \citetalias{2021arXiv210906273R}:
the \hcop{} column density $N_\hcop{}^{\mathrm{obs}}$, the \hi{} optical depth $\tau_\mathrm{HI}^{\mathrm{obs}}$, and the \hi{} spin temperature $T_s^{\mathrm{obs}}$. 
We attempt to find the PDR model that minimizes residual $\chi^2$ between the model and observations by investigating a coarse grid of model parameters, where $\chi^2$ is defined as, 
\begin{equation}\label{eq:chisq}
\begin{split}
    \chi^2
    =& \frac{(N_\hcop{}^{\mathrm{obs}} - N_\hcop{}^\mathrm{PDR})^2}{e^2(N_\hcop{})}
    + \frac{(\tau_\mathrm{HI}^{\mathrm{obs}} - \tau_\mathrm{HI}^\mathrm{PDR})^2}{e^2(\tau_\mathrm{HI})}\\
    &+ \frac{(T_s^{\mathrm{obs}} - T_s^\mathrm{PDR})^2}{e^2(T_s)}.
\end{split}
\end{equation}
This fitting process is not looking for the best-fit model, but a rough estimate of the parameters needed to explain observations.
The uncertainties in the denominators are calculated from
\begin{equation}
    e^2 = e^2_\mathrm{obs} + e^2_\mathrm{sys},
\end{equation}
where $e^2_\mathrm{obs}$ is the observational uncertainty directly obtained from the Gaussian decomposition \citepalias{2021arXiv210906273R}, and $e^2_\mathrm{sys}$ is the systematic uncertainty.
For each slab model at a given $N_\mathrm{H}$, $N_\hcop{}^\mathrm{PDR}$ is obtained directly from the chemical abundances; $\tau_\mathrm{HI}^\mathrm{PDR}$ is the integrated peak optical depth of \hi{} assuming a constant velocity dispersion derived from the observed Gaussian component; and $T_s^\mathrm{PDR}$ is calculated from the average $T_s$ weighted by the \hi{} mass.
The systematic uncertainty $e^2_\mathrm{sys}$ is difficult to estimate. One source of systematic uncertainty is the changes in \hcop{} excitation temperatures, which affect the calculation of $N_\hcop{}$ (see Section \ref{section:comparison_tigress}). Another source of systematic uncertainty lies in the intrinsic assumption of Gaussian decomposition, which presumes that the different parts of the ISM can be represented by uniform structures well-separated in velocity space. \citet{Murray2017} compared the \hi{} observations from 21-SPONGE to numerical simulations, and found a good agreement for \hi{} column density and spin temperature within a factor of $\sim 2$. Here we simply assume a systematic error of 50\% of each observed value. Although $\chi^2$ changes when we vary systematic error fraction, we experimented with varying $e^2_\mathrm{sys}$, and found that the resulting parameters derived from the PDR models are insensitive to the choice of systematic error fraction. For each Gaussian component, we put an upper limit on the total column density $N_\mathrm{H}$ of the slab models to be the $N_\mathrm{H}$ derived from Planck dust emission (Equation \ref{eq:NH}). For sightlines 3C123B and 3C120 where two Gaussian components with \hcop{} detection are identified, the upper limit of column density for each component is simply set to be half of the $N_\mathrm{H}$ derived from dust, as there is no constraint on how much each Gaussian component contributes to the dust column. \footnote{When fitted without any upper limits on $N_\mathrm{H}$, the column densities derived from the two Gaussian components in 3C123B or 3C120 are similar. Therefore, without better knowledge, we choose to simply use half of the dust-based $N_\mathrm{H}$ as the upper limit of column density for each component.}

\subsection{ISM simulations}

While the PDR model provides constraints on the 
underlying physical processes by using 
a one-dimensional view of
the ISM, numerical simulations provide complementary information by enabling a more direct representation of interstellar turbulence taking place in the multi-phase medium.
We use here ISM simulations from \citet{Gong2018} and
briefly describe the key numerical methods. For more detailed descriptions we refer readers to \citet{Gong2018}  and \citet{Gong2020}.

The simulations are taken from the result of R8-Z1 models in \citet{Gong2020}, which represent the solar-neighborhood ISM environment. The simulation resolution is 2 pc and the box-size is 1 kpc in the $x$- and $y$-directions in the galactic disk plane, and 7 kpc in the $z$-direction perpendicular to the galactic disk. The 3D magneto-hydrodynamic simulations are run within the TIGRESS framework \citep{Kim2017, Kim2020}, which models the self-consistent star formation feedback in the three-phase turbulent ISM. In the simulations, gravitational collapse of dense gas leads to the formation of star clusters. Supernova and FUV feedback from the star clusters are calculated from a population synthesis model \citep{Leitherer1999}. The simulations reach a quasi-steady state after about 200 Myr.
The TIGRESS simulations have been widely used for comparison with a large range of ISM observations, such as \hi{} \citep{Murray2017, Murray2020}, CO \citep{Gong2018, Gong2020}, dust polarization \citep{Kim2019} and H$\alpha$ \citep{Kado-Fong2020}. We post-process the outputs to obtain the steady state chemical abundances and gas temperature. Finally, we use the RADMC-3D code \citep{Dullemond2012} to perform radiation transfer calculations to obtain the \hcop{}(1-0) line excitation temperature, adopting the large velocity gradient (LVG) approximation.

Because it is non-trivial to match the structures along the line of sight in simulations to Gaussian components in observed spectra, we simply compare the simulations to the observed total line-of-sight column densities of \hcop{} and \hi{} presented in \citetalias{2021arXiv210906273R}. In the future, we plan to generate synthetic spectra of \hi{} and \hcop{} directly from the simulations, which can be directly compared to observations. 

\section{Comparisons with PDR Models\label{sec:comparision_PDR}}
\begin{deluxetable*}{cccc cccc}
\tablecaption{Gaussian decomposition of the \hcop{} and \hi{} spectra and
the parameters derived from PDR models \label{tab:PDR_model_fit}. The four components measured along sightlines where the \hcop{} absorption was saturated at some velocities are noted with a dagger ($\dagger$). We discuss these in Section \ref{sec:comparision_PDR}.}
\tablehead{
\colhead{Source} &\multicolumn{3}{c}{observed parameters} 
&\multicolumn{4}{c}{parameters derived from PDR models}\\
\colhead{} &\colhead{$N_\hcop{}~(\mathrm{cm^{-2}})$} &\colhead{$\tau_\mathrm{HI}$}
&\colhead{$T_s~(\mathrm{K})$} &\colhead{$n~(\mathrm{cm^{-3}})$} 
&\colhead{$N_\mathrm{H}~(\mathrm{cm^{-2}})$} &\colhead{$G'$} &\colhead{$\xi/\xi_0$}
}
\startdata
3C123B &$(1.74\pm 0.96)\times 10^{12\,\dagger}$ &$0.25\pm 0.01$ &$81.3\pm 43.5$ 
&$7088.0$ &$1.1\times 10^{21}$ &1 &10\\
3C111A &$(1.18\pm 0.02)\times 10^{13\,\dagger}$ &$0.79\pm 0.01$ &$77.1\pm 4.5$ 
&$2848.5$ &$4.8\times 10^{21}$  &100 &10\\
3C123A &$(5.53\pm 1.95)\times 10^{12\,\dagger}$ &$0.81\pm 0.07$ &$67.6\pm 25.6$ 
&$1373.7$ &$3.2\times 10^{21}$  &10 &10\\
3C78 &$(5.90\pm 3.63)\times 10^{10}$ &$0.15\pm 0.00$ &$65.1\pm 3.0$ 
&$10207.0$ &$1.0\times 10^{21}$ &10 &10\\
3C111B &$(3.85\pm 2.54)\times 10^{12\,\dagger}$ &$0.97\pm 0.02$ &$56.1\pm 5.9$
&$552.1$ &$4.2\times 10^{21}$  &10 &1\\
BL Lac &$1.34\pm 0.60 \times 10^{12}$ &$0.22\pm 0.00$ &$41.8\pm 22.9$ &$8505.6$ &$9.8\times 10^{20}$ &1 &10\\
3C123B &$(5.18\pm 2.69)\times 10^{12\,\dagger}$ &$1.65\pm 0.00$ &$36.8\pm 14.1$ &$662.5$ &$1.6\times 10^{21}$  &0.1 &10\\
3C120 &$(5.57 \pm 0.22)\times 10^{10}$ &$0.78\pm 0.01$ &$23.9\pm 5.1$ &$128.4$ &$6.6\times 10^{20}$  &0.1 &1\\
3C120 &$(2.80\pm 0.15)\times 10^{11}$ &$0.62\pm 0.01$ &$20.5\pm 9.8$ &$460.1$ &$8.8\times 10^{20}$  &0.1 &1\\
3C454.3 &$(2.98\pm 0.00)\times 10^{11}$ &$0.31\pm 0.02$ &$20.1\pm 2.0$ &$552.1$ &$7.5\times 10^{20}$  &0.1 &1\\
3C154 &$5.43\pm 0.68 \times 10^{11}$ &$1.30\pm 0.01$ &$10.4\pm 15.1$ &$61.9$ &$2.1\times 10^{21}$  &0.1 &0.1\\
\enddata
\end{deluxetable*}

\begin{figure}[] 
\gridline{\fig{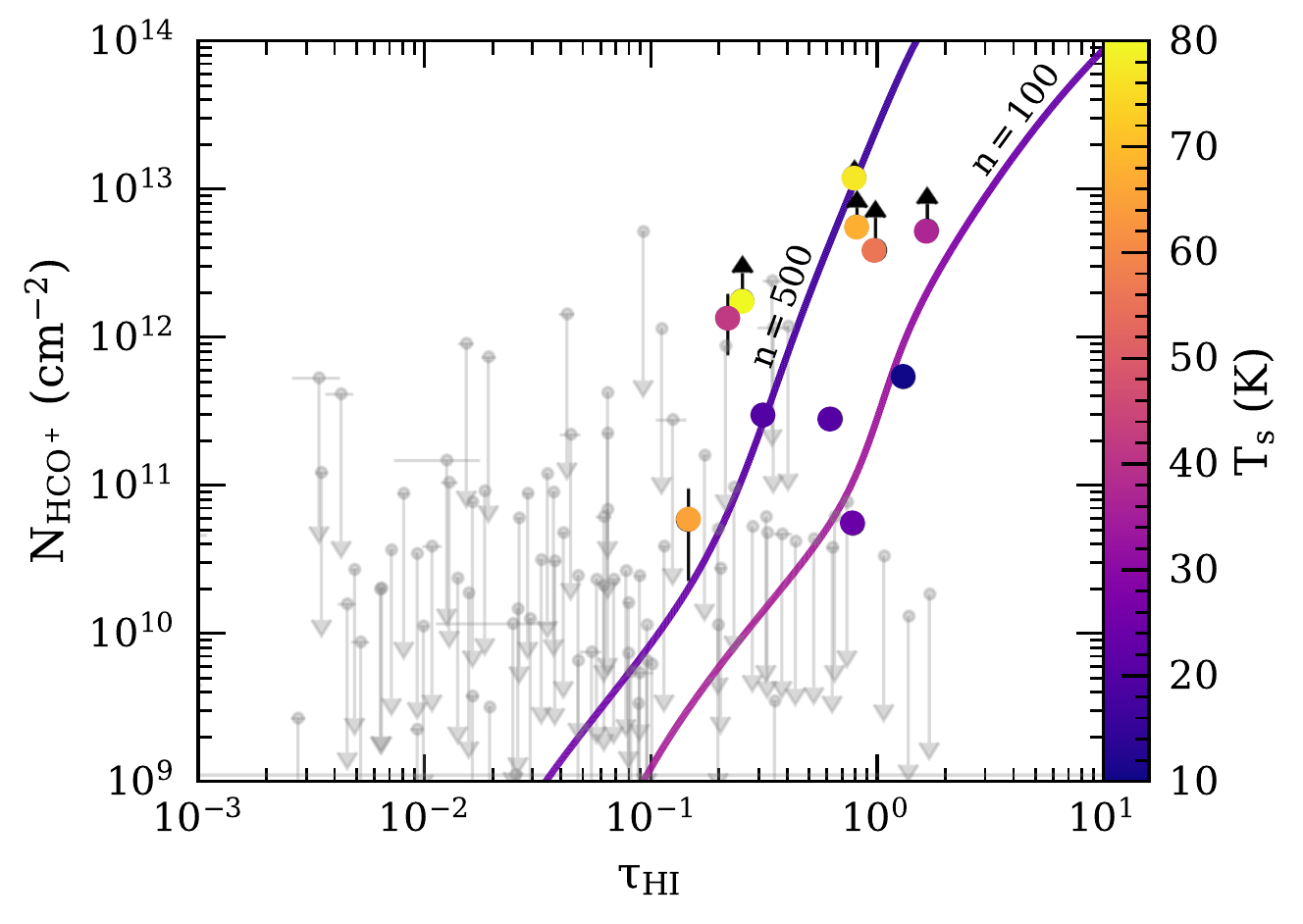}{\columnwidth}{$G'=0.1,~ \xi=2\times 10^{-16}s^{-1}$}}
\gridline{\fig{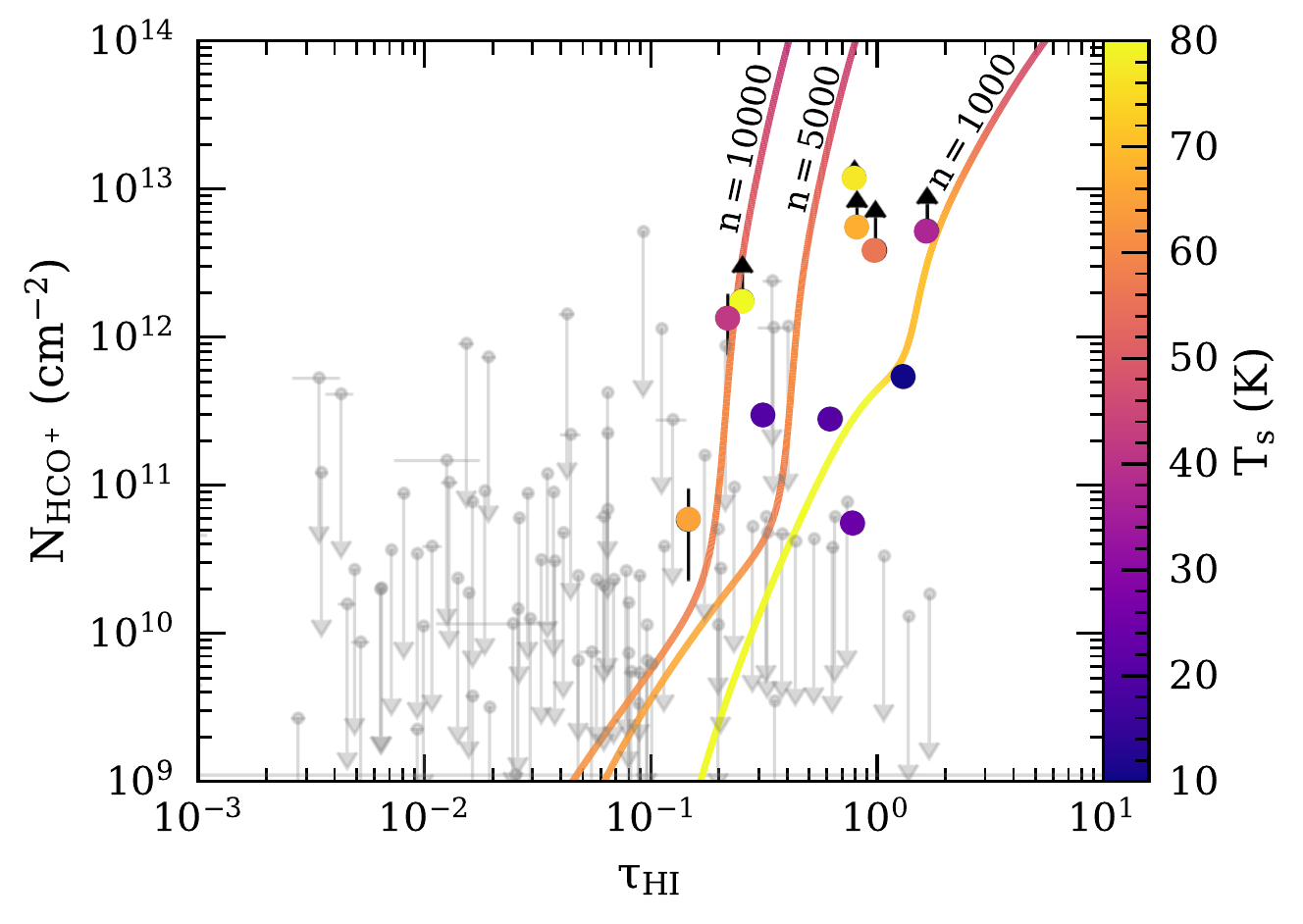}{\columnwidth}{$G'=10,~ \xi=2\times 10^{-15}s^{-1}$}}
\caption{Comparison between the 1D PDR models (curves) in \citet{Gong2017} and the observation results from Gaussian decomposition (circles, see Table \ref{tab:PDR_model_fit}). The x-axis shows the \hi{} optical depth, the y-axis the \hcop{} column density, and the color scale the \hi{} spin temperature. The optical depth is calculated assuming a constant velocity dispersion of 1 \kms{} in the models (the typical FWHM in each Gaussian component is 1.1--4.4 \kms{}). Different curves shows the PDR models with different volume densities $n$ labeled in the plot (in units of $\mathrm{cm^{-3}}$ for H atom). Each point on the curve correspond to a different column density in the PDR, colored by the mass-weighted \hi{} spin temperature throughout this column. The gray points shows the upper limits in the observations. 
The upper panel shows the PDR model with low incident FUV radiation field ($G'=0.1$) and low CRIR ($\xi=2\times 10^{-16}s^{-1}$), and the lower panel shows the model with high incident FUV radiation field ($G'=10$) and high CRIR ($\xi=2\times 10^{-15}s^{-1}$). The data points with low $T_s$ are more consistent with the low FUV radiation and CRIR environment at low gas density, while the data points with high $T_s$ are more consistent with the high FUV radiation and CRIR environment at high gas density (see also Table \ref{tab:PDR_model_fit}). Note that the uncertainty in $T_s$ is now shown in the plot, but can be read from Table \ref{tab:PDR_model_fit}. 
}
\label{fig:NHCOp_tauHI}
\end{figure}

The parameters for the most reasonable PDR models estimated for each
Gaussian feature detected in \hcop{} and \hi{} absorption are given in Table \ref{tab:PDR_model_fit}. Even with a very coarse grid of $G'$ and $\xi$, the PDR model can fit the observed Gaussian components reasonably well, with the residual $\chi^2\lesssim 1$ 
(Equation \ref{eq:chisq}). 

Table \ref{tab:PDR_model_fit} is arranged with descending values of $T_s$, and one can immediately see that there are two distinct groups in the parameters derived from PDR models.
When $T_s \gtrsim 40~\mathrm{K}$, the data points are best reproduced with high density, $n\approx 500 - 10^4~\mathrm{cm^{-3}}$, and higher values of $G'\gtrsim 1$ and $\xi/\xi_0 \gtrsim 1$. These dense PDRs exposed to strong FUV and CR radiation are indicative of environments close to the formation of massive stars, such as the edge of \hii{} regions \citep[][see Section \ref{subsec:PDR_fits_v_observations} for more discussion]{HT1999}.
All sightlines in this group also have a significant fraction of thermally unstable \hi{} \citepalias[UNM; see][for details]{2021arXiv210906273R}.
Contrarily, when $T_s \lesssim 40~\mathrm{K}$, the data points are best reproduced by models with lower density $n\approx 60-700~\mathrm{cm^{-3}}$ and lower values of $G'\lesssim 1$ and $\xi/\xi_0 \lesssim 1$. 
Based on our \citetalias{2021arXiv210906273R}, three out of five of these components are in directions with negligible thermally unstable HI.
These environments are consistent with classic diffuse molecular gas in the Solar neighborhood \citep{Draine2011, Tielens2013}. 

We note that our grid of models is very coarse, and there are significant uncertainties with regard to each parameter. For example, the parameters derived from PDR models change somewhat when we use two-sided instead of one-sided slab models, when we increase $N_\hcop{}$ by a factor of 2 considering the effect of line saturation (Table \ref{tab:PDR_model_fit}), when we consider the likely under-estimation of excitation temperature (Figure \ref{fig:Texc_sim}), or when we do not impose an upper limit on the total column density $N_\mathrm{H}$ considering the dust-based $N_\mathrm{H}$.
However, in each of these cases, we still see two distinct groups: one with high $T_s \gtrsim 40~\mathrm{K}$, high FUV radiation $G'\gtrsim 1$, high CRIR $\xi/\xi_0 \gtrsim 1$ and high density $n \gtrsim 1000~\mathrm{cm^{-3}}$; and another with low $T_s \lesssim 40~\mathrm{K}$, low FUV radiation $G'\lesssim 1$, low CRIR $\xi/\xi_0 \lesssim 1$ and low density $n \lesssim 1000~\mathrm{cm^{-3}}$. In short, the detailed parameters 
from our selected PDR
models are likely to have large uncertainties, but the conclusion that the PDR model fitting results in these two distinct groups of data points separated by $T_s$ is robust. This clearly demonstrates the power of using \hi{} to characterize the underlying physical conditions where molecules form and reside.

Figure \ref{fig:NHCOp_tauHI} further demonstrates the behavior of these two groups. In order to achieve higher $T_s$, stronger FUV radiation and higher CRIR are needed to heat up the gas. However, FUV radiation and CRs also lead to the destruction of molecular gas, and thus higher densities are needed to form the observed column of \hcop{}. We note that the fractional abundance of \hcop{} does not saturate until $\tau_\mathrm{HI} \gtrsim 1$. Therefore, in the bottom panel of Figure \ref{fig:NHCOp_tauHI}, the curves show that $N(\hcop{})$ steeply increases with $\tau_\mathrm{HI}$ as the \hcop{} abundance increases. In the top panel, the $N(\hcop{})$--$\tau_\mathrm{HI}$ relation is close to linear at $\tau_\mathrm{HI} \gtrsim 1$, as the \hcop{} abundance approaches a constant value.

Previously, \citet{2010A&A...520A..20G} found that PDR models underpredicted the observed \hcop{} column densities by an order of magnitude for many diffuse lines of sight. They used the Meudon PDR code \citep{2006ApJS..164..506L} with $n=10^2\text{--}10^4$ \percc{}, $G^\prime=1\text{--}3$, and $\xi/\xi_0=0.15$. 
Here we show that PDR models can reproduce the high observed \hcop{} abundances in the diffuse ISM (as well as environmental constraints provided by \hi{}), but only when a much broader range of FUV radiation fields and CRIRs is allowed. 
In particular, the CRIR is significantly different between our models and those used in \citet{2010A&A...520A..20G}.
They used $\zeta_0=3 \times 10^{-17}$ s$^{-1}$, i.e., $\xi/\xi_0=0.15$, for all models.
Observations of H$_3^+$, which due to its simple chemistry is one of the most direct tracers of the CRIR, have indicated a mean CRIR of a few times $10^{-16}$ s$^{-1}$  in the Milky Way's diffuse ISM \citep{McCall2003,Indriolo2007}, with measurements spanning an order of magnitude, from $\sim0.5\times10^{-16}$ s$^{-1}$ to $\sim5\times10^{-16}$ s$^{-1}$ \citep{vanderTak2006,Indriolo2007,Indriolo2012}. These observationally-derived CRIRs tend to be much higher than the canonical value used in \citet{2010A&A...520A..20G}.
The models used in this work range from $\xi\approx0.1\xi_0$  \citep[similar to the value used in][]{2010A&A...520A..20G}, to $\xi\approx10\xi_0$.
As discussed in \citet{2010A&A...520A..20G}, the CR ionization rate has a strong influence on the \hcop{} column density, independent of density.

In Section \ref{subsec:PDR_fits_v_observations} we further discuss whether the broad ranges in $n$, $G^\prime$, and $\xi_0$ required by the PDR to approximate the observed \hcop{} and \hi{} gas properties (Table \ref{tab:PDR_model_fit}) are plausible. We argue that environments where the PDR model requirements are incompatible with observations likely represent sites of non-equilibrium chemistry not represented in the PDR models.

\section{Comparisons with Multi-phase ISM simulations\label{section:comparison_tigress}}

\begin{figure}[] 
\gridline{\fig{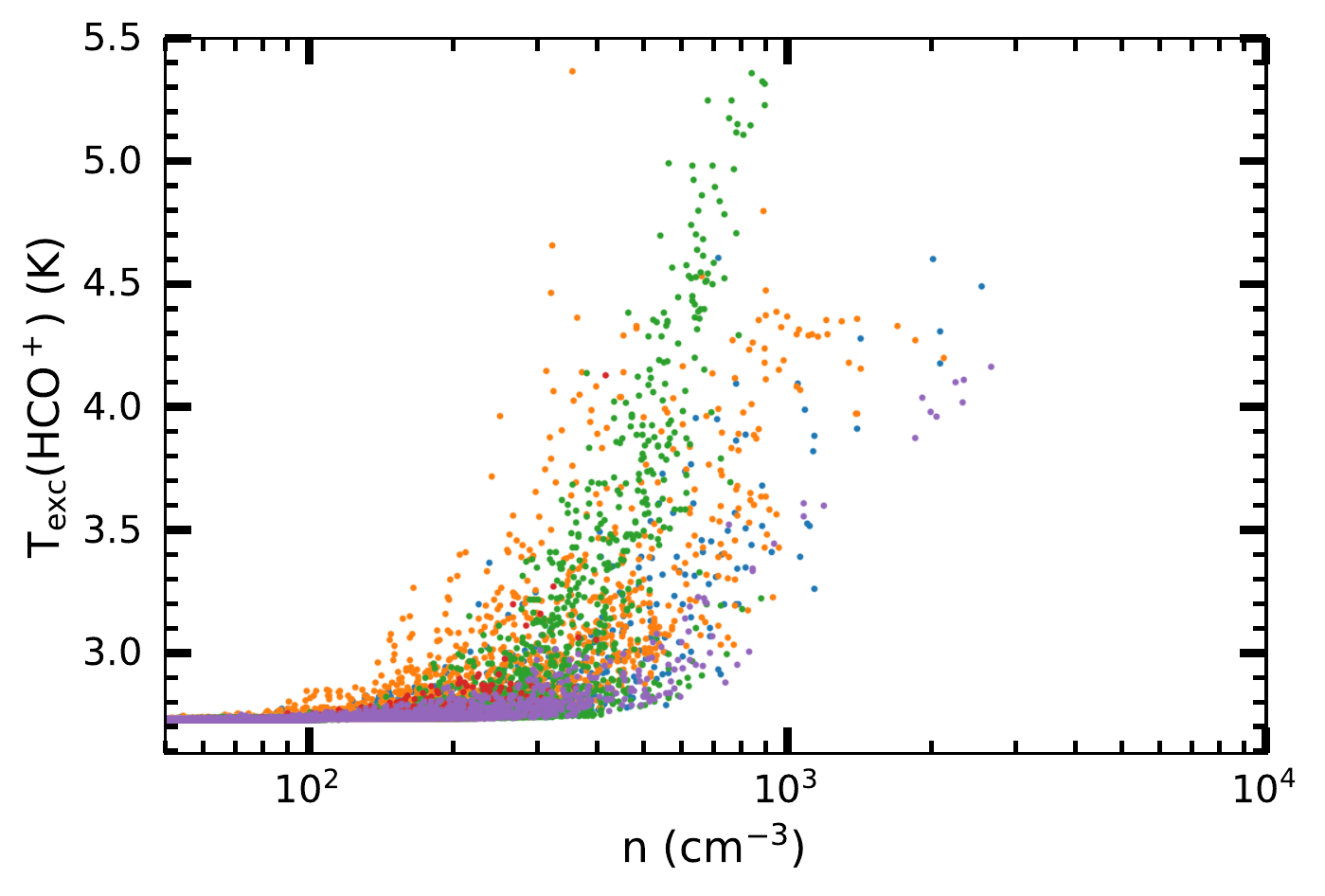}{\columnwidth}{}}
\caption{\hcop{}(1-0) excitation temperature $T_\mathrm{exc}(\hcop{})$ versus gas density $n$ in the numerical simulations in \citet{Gong2020}.
Each scattered point represent a grid cell the size of $(2\mathrm{pc})^3$, and the different colors represent different snapshots in time, where the gas density distribution, ambient FUV radiation field vary. 
At $n\gtrsim 300~\mathrm{cm^{-3}}$, the \hcop{} excitation temperature rises above the CMB background temperature.
\label{fig:Texc_sim}}
\end{figure}

\begin{figure*}[]
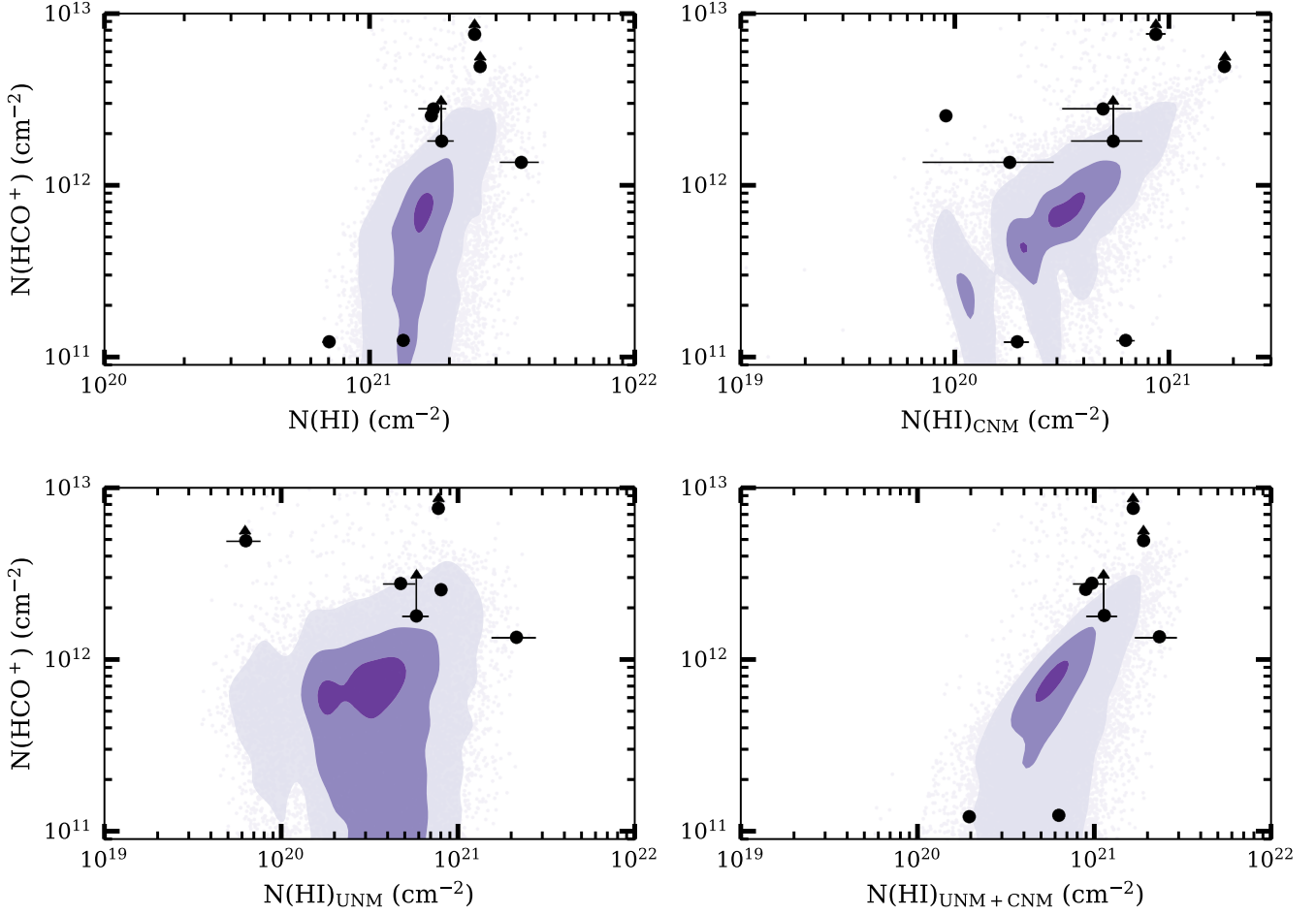
 
\gridline{\fig{NHCOP_NH_sim.pdf}{\textwidth}{}}
\caption{Comparison between the numerical simulation in \citet{Gong2020} (contours) and the observed \hcop{} and \hi{} column densities (scattered points).
The contours shows the 10\%, 50\%, and 90\% density levels of the pixels with $N(\hcop{}) > 10^{10}~\mathrm{cm^{-2}}$, viewing along the z-axis of the simulation (face-on).  Two sightlines (3C120 and 3C454.3) have no \hi{} in the UNM, so only six points appear in the lower left panel.
\label{fig:NHCOP_NH_sim}}
\end{figure*}

The MHD simulations allow us to investigate a common assumption used in calculating the \hcop{} column density. In Section \ref{subsec:hcop_obs} we assumed, like many previous studies, that the excitation temperature is equal to the CMB temperature 
\citep[e.g.,][]{2010A&A...520A..20G,2020ApJ...889L...4L}. However, we can check this assumption using the simulated data. In the ISM simulation, the level population of \hcop{} is calculated by RADMC-3D code using the LVG approximation, where the escape probability of the photon is determined by the local velocity gradient. A background black-body radiation field of 2.725 K is included, as well as collisional excitation by the \htwo{} molecule. The level population is solved iteratively, from which we derive the excitation temperature.

The excitation temperature of \hcop{} versus gas density is shown in Figure \ref{fig:Texc_sim}. At density $n\gtrsim 300~\mathrm{cm^{-3}}$, the excitation temperature $T_\mathrm{exc}(\hcop{})$ rises above the CMB temperature. 
This clearly shows that 
the observed $N(\hcop{})$ can be under-estimated in the higher density regions. For example, if the excitation temperature of \hcop{} were 5 K (Figure \ref{fig:Texc_sim}), then the estimated \hcop{} column densities would be a factor of $2.2$ too low. There are also significant variations in $T_\mathrm{exc}(\hcop{})$ across snapshots in the simulations, where gas structure, FUV radiation field strength, and CRIR differs. This indicates that directions with higher $T_s$ in Table 1 likely have under-estimated \hcop{} column densities. This will affect the comparison with the PDR model in Figure \ref{fig:NHCOp_tauHI}.

In Figure \ref{fig:NHCOP_NH_sim} we compare the total line-of-sight column densities of \hcop{} and \hi{}, both from the simulation and from the observations. The total \hi{} column density is shown in the upper left panel. We can track in the simulation \hi{} in different phases.
The CNM column density (the column density of \hi{} with $T_s<250~\rm{K}$) is shown in the upper right panel. The UNM column density (the column density of \hi{} with $250~\mathrm{K} < T_s < 1000~\mathrm{K}$) is shown in the lower left panel. The sum of the CNM and UNM column densities is shown in the lower right panel. This is similar to Figure 3 of \citetalias{2021arXiv210906273R}. We plot one representative time-snapshot in the simulation, as we did not find a large variation with time. The line-of-sight column density in the simulations is calculated along the $z$-axis (face-on), and we found that viewing along the $x$- or $y$- axis (edge on) gives similar results. We note that only pixels with $N(\hcop{}) > 10^{10}~\mathrm{cm^{-2}}$ are shown in Figure \ref{fig:NHCOP_NH_sim}. In fact, most of the sightlines in the simulations only have the WNM component with negligible  amount of \hcop{}.

Both the simulation and observations show a transition from an atomic to molecular region at $N(\hi{})\approx 10^{21}~\mathrm{cm^{-2}}$. This is mainly due to the \htwo{} self-shielding (with a small contribution from dust-shielding) at $A_V\lesssim 1$, which attenuates the FUV radiation and allows molecules to form in the absence of fast photo-dissociation. 

Similarly, both the observations and simulation show that \hcop{} is detected only where the column density of \hi{} in the CNM is $\gtrsim10^{20}$ \persc{}. The lower right panel further shows that the total amount of cold \hi{} (CNM plus UNM) is correlated with the \hcop{} column density for $N(\hi{})_{\rm{CNM+UNM}}\gtrsim2\times10^{20}$ \persc{}, below which no \hcop{} is detected.

There are a few features that indicate potential tension between the observations and simulation, though. First, there are two distinct groups in the observations with low $N(\hcop{}) \approx 10^{11}~\mathrm{cm^{-3}}$ and high $N(\hcop{}) > 10^{12}~\mathrm{cm^{-3}}$, while the simulation shows a continuous distribution with most sightlines having intermediate values of  $N(\hcop{})$. Second, a couple of observed sightlines show a very high value of $N(\hcop{}) \gtrsim 2\times10^{12}~\mathrm{cm^{-3}}$, which is very rare in the simulations. One possible cause is the fact that we did not include \hii{} regions in the simulations, which can produce ionisation fronts with high gas density and incident radiation \citep{HT1999} As discussed in Section \ref{sec:comparision_PDR}, dense PDRs close to \hii{} regions can be the site for \hcop{} formation. In fact, sources 3C111A, 3C123A and 3C123B have both total $N(\hcop{}) \gtrsim 2\times10^{12}~\mathrm{cm^{-3}}$ as well as Gaussian decomposition components of high \hi{} spin temperature and \hcop{} column. 
It will be very interesting to investigate if formation of \hcop{} near \hii{} regions can explain such high values of $N(\hcop{})$.
Finally, almost all simulated data have a significant fraction of UNM, and the lower left panel of Figure \ref{fig:NHCOP_NH_sim} suggests that a UNM column density threshold of $10^{20}$ \persc{} is required for \hcop{} formation. As we discussed in \citetalias{2021arXiv210906273R}, though, $40\%$ of observed directions in our sample have no detectable UNM, including 2 of the 9 sightlines that showed \hcop{} absorption (3C120 and 3C454.3).

Currently, we are working on a new set of numerical simulations with radiation feedback and \hii{} regions from massive stars, as well as more accurate modeling of gas heating and cooling. In the future, direct comparisons between the observed spectra and simulated synthetic spectra, as well as more observational data, will put better constraints on the formation of molecules in the ISM.

\section{Discussion}
\label{sec:discussion}

\subsection{Are the PDR parameters realistic?}
\label{subsec:PDR_fits_v_observations}

While results in Section \ref{sec:comparision_PDR} suggest that the \hcop{} column densities reported in \citetalias{2021arXiv210906273R} may be underestimated in a few directions by a factor of a few due to the assumption of a constant (CMB) excitation temperature (and an even larger factor for 3C111A, 3C111B, and 3C123B due to absorption line saturation), we investigate here whether environmental estimates given in Table \ref{tab:PDR_model_fit} are reasonable.

We can provide a rough  estimate of the mean UV radiation field along a line of sight using the dust temperature $T_d$ and power law index $\beta$, namely $G^\prime \propto T_d^{\beta+4}$ (Section \ref{subsec:planck}). Taking $T_d$ and $\beta$ from \textit{Planck} \citep{2014A&A...571A..11P,2016A&A...596A.109P}, we find that $G^\prime \sim 0.52$ for 3C123, $G^\prime \sim 0.44$ for 3C111, $G^\prime \sim 0.90$ for 3C78, and $G^\prime \sim 0.90$ for BL Lac. These values are essentially similar to the Solar neighbourhood ambient radiation field and are in agreement with PDR predictions for 3C123B, BLLac, 3C120, 3C454.3, and 3C154, but are much lower than PDR predictions for 3C111A, 3C123A, 3C78 and 3C111B. There is no systematic difference in $T_d$ or the estimated FUV radiation field in the direction of the sources with high $T_s$ structures in Table \ref{tab:PDR_model_fit} from those with low $T_s$ structures. 
We note that the \textit{Planck} maps provide only line-of-sight integrated properties, while from \hi{} we know that there are multiple absorption structures along each line of sight listed in Table \ref{tab:PDR_model_fit}, as well as multiple molecular absorption structures along all but 3C78. It is also known that $G^\prime$ can vary on vary small spatial scales, especially near O stars \citep[e.g.,][]{Marconi1998}. Nevertheless, the star formation rates in California and Taurus are modest \citep{2010ApJ...724..687L}, and neither 3C111, 3C123, nor 3C78 are particularly close to any known \hii{} regions or protostellar objects \citep{2017A&A...606A.100L}. While it is difficult to definitively rule out the values of $G^\prime$ given in Table \ref{tab:PDR_model_fit} based on the above discussion, it is likely that PDR model parameters for 3C111A, 3C123A, 3C78 and 3C111B---an order of magnitude greater than the local mean---are too high.

Estimates of the local density in the diffuse ISM along our exact lines of sight are also difficult to obtain. In general, estimates of the densities of diffuse molecular gas structures are modest \citep[$\sim$a few $\times10$ \persc{} to a few $\times10^2$ \percc{};][]{Goldsmith2013}, consistent with estimates from the PDR model for structures with $T_s<40$ K. 3C111 is behind the California molecular cloud, and 3C111 and 3C123 both probe gas from the Taurus molecular cloud, so we can compare the estimates in Table \ref{tab:PDR_model_fit} to estimates of the densities in these well studied structures. Previous works have estimated the densities of cores \citep[$\sim10^4$ \percc{}, assuming $G\sim0.1\text{--}1$;][]{2010ApJ...721..686P} and filaments \citep[$10^3$--$10^4$ \percc{};][]{2021A&A...648A.120R} in Taurus and  clumps in the direction of California \citep[$10^2$--$10^3$ \percc{};][]{1991A&A...249..483H}. The maximum densities toward 3C123 and 3C111 in Table \ref{tab:PDR_model_fit} are $10^3$--$10^4$ \percc{}, which are similar in magnitude to those of cores and filaments in Taurus and California. However, extinction maps derived from \textit{Planck} \citep[full sky, $5\arcmin$ resolution][]{2014A&A...571A..11P}, from point sources in the Two Micron All Sky Survey \citep[Taurus and California, $2.5\arcmin$ resolution][]{2010A&A...512A..67L}, and from \textit{Herschel} \citep[California, $1.3\arcmin$ resolution][]{2017A&A...606A.100L} do not show exceptionally high extinction toward 3C111 or 3C123 (observations of cores, clumps, and filaments had higher resolution, $12$--$47\arcsec$). 
Nevertheless, \citet{2014ApJ...784..129A} established rough limits on the gas densities in the direction of 3C111 (for the gas structures between $-5$ \kms{} and $+2$ \kms{}, the same as those listed in Table \ref{tab:PDR_model_fit}) in the range of $10^3$--$10^5$ \percc{} from observations of H$_2$CO. The lower limit was established by comparison of CO and \hcop{} line profiles and the upper limit was established by the low observed excitation temperatures of H$_2$CO.
A wide variety of other molecular lines has also been detected in absorption in the direction of 3C111, including HCO, H\textsuperscript{13}CO\textsuperscript{+}, c-C\textsubscript{3}H \citep{2014A&A...564A..64L}, OH \citep{1996A&A...314..917L}, \textsuperscript{12}CO, \textsuperscript{13}CO, C\textsuperscript{18}O \citep{1998A&A...339..561L}, \cch{}, C\textsubscript{3}H\textsubscript{2} \citep{2000A&A...358.1069L}, CN, \hcn{}, \hnc{} \citep{2001A&A...370..576L}, CS, SO, H\textsubscript{2}S, HCS\textsuperscript{+} \citep{2002A&A...384.1054L}, CH \citep{2002A&A...391..693L}, $l$-C\textsubscript{3}H, HC\textsubscript{3}N, and CH\textsubscript{3}CN \citep{2018ApJ...856..151L}. 
The rich diversity of molecular species identified in this direction testifies to a complex chemistry, and likely implies high density. 
Observations toward 3C123 have been rarer, although \citet{2018ApJ...862...49N} recently identified three OH absorption components in the direction of 3C123, all consistent with the features we find at $\sim3.3$ \kms{}, $\sim4.3$ \kms{}, and $\sim5.3$ \kms{}. OH is a diffuse molecular gas tracer \citep{Li2018}, so it is not clear based on existing observations if a density of $10^3$ \percc{} (Table \ref{tab:PDR_model_fit}) is plausible.

The two highest density estimates are in the direction of 3C78 and BL Lac, with $n\sim10^4$ \percc{}. The sightline toward 3C78 probes the outer layers of the starless molecular cloud  MBM16 \citep{MBM1985} and may be directly probing a separate molecular cloud identified by \citet{LaRosa1999}. H$_2$CO observations in the direction of MBM16 suggest a density $\lesssim10^3$ \percc{} \citep{1993ApJ...402..226M}, but the 3C78 sightline only probes the outermost layers of MBM16. The density of the second molecular cloud is not known. \citet{LaRosa1999} detected CO in this direction, but we do not detect significant CO emission from \citet{2001ApJ...547..792D} in the pixels nearest 3C78, nor do we see \cch{}, \hcn{}, or \hnc{} absorption in this direction, probably indicating that a density of $10^4$ \percc{} is not plausible. Meanwhile, BL Lac probes the Lacerta molecular cloud. As with 3C111, \citet{2014ApJ...784..129A} inferred a density of $\sim10^3$--$10^5$ \percc{} in this direction from the properties of H$_2$CO absorption. The result in Table \ref{tab:PDR_model_fit}, $n\approx10^4$ \percc{}, is consistent with these bounds.

The CRIR is not observationally well constrained along our lines of sight. It is known that the CRIR does vary between different lines of sight by as much as an order of magnitude, even in the diffuse ISM \citep{vanderTak2006,Indriolo2007,Indriolo2012}. If low energy cosmic rays are accelerated in localized shocks, it is expected that sightlines near more energetic regions, such as regions of massive star formation, will have higher CRIRs \citep{Indriolo2012}, although this has not been tested observationally. CRIRs measured from observations of diffuse molecular gas have been as high as $\xi\approx10\xi_0$ \citep{2008ApJ...675..405S}, which is consistent with the higher estimates in Table \ref{tab:PDR_model_fit}.

In summary, the \hcop{} column densities of structures with $T_s<40$ K are well explained by PDR models with modest densities, FUV fields, and CRIRs, consistent with what is expected for local diffuse molecular gas. On the other hand, the \hcop{} column densities of structures with $T_s>40$ K are only explained by PDR models with systematically higher densities, FUV fields, and CRIRs. Previous observations in the direction of these structures suggest that the high densities in Table \ref{tab:PDR_model_fit} may be plausible for some structures (3C111, BL Lac), but are very likely too high in at least one case (3C78). We also do not find evidence for extremely strong FUV fields in any direction, nor are the star formation rates particularly high for molecular clouds in these directions like California or Taurus. This suggests that the estimates of the FUV field in Table \ref{tab:PDR_model_fit} may also be too high for some of these structures.

\subsection{Equilibrium versus non-equilibrium chemistry in the diffuse ISM}
The \citet{Gong2017} PDR (UV-dominated, equilibrium) chemical models appear sufficient to explain the observed \hcop{} column densities for the structures with $T_s<40$ K along lines of sight that tended to have lower fractions of thermally unstable \hi{}, without the need for non-equilibrium chemistry or nearby \hii{} regions. While the densities ($\sim60$--$700$ \percc{}) and FUV field strengths ($G^\prime\approx0.1$) for these features were moderate, we note that in most cases the CRIR was  $\xi\approx\xi_0$. This is consistent with the observationally derived CRIR from H$_3^+$ \citep{vanderTak2006,Indriolo2007,Indriolo2012}, but about an order of magnitude higher than the CRIR used in the PDR models in \citet{2010A&A...520A..20G} \citep[$3\times10^{-17}$ $\rm{s^{-1}}$;][]{Dalgarno2006}. As noted by \citet{2010A&A...520A..20G}, increasing $\xi$ by an order of magnitude increases the column density of \hcop{} by an order of magnitude, regardless of gas density.  This suggests that the CRIR may be partly responsible for the discrepancy between their PDR model predictions and \hcop{} observations (although we not that this is not the case for all species).

Meanwhile, structures with $T_s>40$ K that were found along sightlines with high fractions of thermally unstable \hi{} were most consistent with PDR models that had very high density, FUV field, and CRIR. Based on the discussion in the previous section, the enhancement of \hcop{} near \hii{} regions is probably not responsible for the high \hcop{} column densities along these lines of sight. In at least one case (3C78), we can definitively reject several of the parameters in Table \ref{tab:PDR_model_fit}, likely indicating that non-equilibrium effects play a role in producing \hcop{} in this direction.
In \citetalias{2021arXiv210906273R}, we showed that several of these structures had higher turbulent velocities ($v_t\gtrsim1$ \kms{} for the structures with $T_s > 40$ K and $v_t\lesssim1$ \kms{} for the structures with $T_s < 40$ K), suggesting they probe more turbulent environments than the low $T_s$ structures.
Moreover, the sightlines with high $T_s$ structures in Table \ref{tab:PDR_model_fit} also tend to have a high fraction of thermally unstable \hi{}. Interestingly, the UNM is thought to be enhanced by the turbulent mixing of the CNM and UNM \citep[e.g.,][]{2005A&A...433....1A,2014A&A...567A..16S}---a process also thought to enhance molecular abundances \citep{Lesaffre2007}. The extent to which the UNM is a tracer of turbulent mixing or turbulent dissipation, though, depends on assumptions about the fraction of turbulent energy injected into each atomic gas phase. Nevertheless, the incompatibility of the PDR model parameters and existing observations in at least one case (and perhaps several cases), coupled with the large quantity of thermally unstable gas along these sightlines, argues in favor of alternative, non-equilibrium chemical models in these directions.

\subsection{Future work}

In the future, observations of multiple molecular transitions can be used to more tightly constrain the environment of molecular clouds along these lines of sight. 
Observations of a wider range of chemical species may also be used as a probe for non-equilibrium chemistry to elucidate the role of turbulent dissipation, shocks, and other dynamical processes in diffuse interstellar chemistry. For example, \citet{2018ApJ...862...49N} measured three OH absorption features in the direction of 3C154 at $-2.32$ \kms{}, $-1.39$ \kms{}, and $2.23$ \kms{}. In \citetalias{2021arXiv210906273R}, we detected strong absorption for \cch{}, \hcn{}, \hcop{}, and \cch{} at $-2$\kms{} and $-1.3$ \kms{}. Additional, somewhat broader components were also identified in the \hcop{}, \cch{}, and \hcn{} absorption spectra at $2.3$ \kms{} and $-4$ \kms{}. The latter component was not detected in OH absorption by  \citet{2018ApJ...862...49N}, possibly suggesting that \hcop{}, \cch{}, and \hcn{} are enhanced relative to OH in this particular structure. The TDR model predicts that at high \hcop{} column density, these species are enhanced relative to PDR predictions, while OH abundances are depressed \citep{2009A&A...495..847G}. This feature is not included in Table \ref{tab:PDR_model_fit} because $T_s$ is unknown for the \hi{} structure \citep{2018ApJS..238...14M}. Nevertheless, this hints at non-equilibrium chemistry and emphasizes the importance of observing multiple species to disentangle PDR and TDR predictions.
Species like CH\textsuperscript{+} are also enhanced in TDRs, and SiO and HNCO are shock tracers. 
A comparison of observed absorption line profiles to simulated synthetic spectra may also help identify regions where chemistry is out of equilibrium---it remains unclear, for example, if the broad component of \hcop{} identified in the ALMA-SPONGE and NOEMA-SPONGE spectra \citepalias[see][]{2021arXiv210906273R} can be explained by PDR models or if this spectral signature is indicative of turbulent dissipation \citep[e.g.,][]{2006A&A...452..511F}.

\section{Conclusion} \label{sec:conclusions}
We compare predictions from the \citet{Gong2017} PDR chemical model and the \citet{Gong2020} ISM simulations 
with \hcop{} observations from \citetalias{2021arXiv210906273R} and \hi{} observations from 21-SPONGE \citep{2015ApJ...804...89M,2018ApJS..238...14M}. It has previously been observed that \hcop{} column densities are generally higher than predicted from PDR models with solar neighborhood conditions \citep[e.g.,][]{2010A&A...520A..20G}. 
Here, by using \hi{} observations, we are able to diagnose interstellar conditions in which high column densities of \hcop{} are found.
Using a coarse grid of PDR models, we estimate the density, FUV radiation field, and CRIR for each structure we identify in \hcop{} absorption. 
We find that absorbing structures fall into two categories---warmer  features ($T_s\gtrsim40$ K) mostly with high \hcop{} column densities ($\gtrsim10^{12}$ \persc{}) that are best reproduced with high density, FUV radiation field, and CRIR models ($n\sim10^3$--$10^4$ \percc{}, $G^\prime\gtrsim1$, $\xi/\xi_0\gtrsim1$); and cooler features ($T_s\lesssim40$ K) mostly with low \hcop{} column densities ($\lesssim10^{12}$ \persc{}) that are best reproduced by low density, FUV radiation field, and CRIR models ($n\sim10^2$--$10^3$ \percc{}, $G^\prime\lesssim1$, $\xi/\xi_0\lesssim1$). The latter are typical of the diffuse molecular ISM in the solar neighborhood, whereas the former are more characteristic of environments close to the formation of massive stars. 
The important diagnostics provided by \hi{} observations demonstrated that high \hcop{} column density features have systematically higher $T_s$ than the features with lower \hcop{} column densities, and they are also found along lines of sight with higher fractions of thermally unstable \hi{} \citep[$30$--$70\%$][]{2018ApJS..238...14M} than the lower column density features, a majority of which have negligible amounts thermally unstable \hi{}.

In at least one case, the parameters derived from PDR models for a structure with $T_s>40$ K is at odds with existing observations. For 3C78, the estimated density of $10^4$ \percc{} and FUV field of $G^\prime\sim10$ are implausible given that \cch{}, \hcn{}, and \hnc{} are not detected in this direction and that there are no nearby sites of massive star formation. The densities and FUV fields required by the PDR model for several other structures, all with $T_s>40$ K, are also likely too high. A likely alternative explanation for the high \hcop{} column densities in these environments is \hcop{} production through dynamical processes such as shocks or turbulent dissipation \citep{2009A&A...495..847G,2010A&A...520A..20G,Valdivia2017}. Future observations (including, e.g., direct observations of shock tracers, isotopologues of CO and \hcop{} that are not saturated, and species whose predicted line ratios are highly model dependent) are needed to distinguish between different non-equilibrium chemical models.

Finally, we show that observations and simulations of the turbulent, multi-phase ISM \citep{Gong2020} agree that \hcop{} formation occurs when the total hydrogen column density reaches $\sim10^{21}$ \persc{}. However, the simulated data fail to explain \hcop{} column densities $\gtrsim ~ \rm{few}\times10^{12}$ \persc{}. Since six of the nine sightlines with detected \hcop{} absorption from \citetalias{2021arXiv210906273R} had such high column densities, this may again indicate that non-equilibrium chemistry is important for these lines of sight.

\acknowledgements{
Support for this work was provided by the NSF through award SOSPA6-023 from the NRAO.
S.S. acknowledges the support by the Vilas funding provided by  the  University  of  Wisconsin  and  the  John  Simon Guggenheim fellowship.
M. G acknowledges the support by Paola Caselli and the Max Planck Institute for Extraterrestrial Physics.
This work is based on observations carried out under project number W19AQ and S20AB with the IRAM NOEMA Interferometer.  IRAM is supported by INSU/CNRS (France), MPG (Germany) and IGN (Spain).
This paper makes use of the following ALMA data: ADS/JAO.ALMA\#2018.1.00585.S and ADS/JAO.ALMA\#2019.1.01809.S.. ALMA is a partnership of ESO (representing its member states), NSF (USA) and NINS (Japan), together with NRC (Canada), MOST and ASIAA (Taiwan), and KASI (Republic of Korea), in cooperation with the Republic of Chile. The Joint ALMA Observatory is operated by ESO, AUI/NRAO and NAOJ. The National Radio Astronomy Observatory is a facility of the National Science Foundation operated under cooperative agreement by Associated Universities, Inc.

\software{PDR code \citep{Gong2017}, RADMC-3D \citep{Dullemond2012}, dustmaps \citep{2018JOSS....3..695M}, CASA \citep{2007ASPC..376..127M}}.
}

\bibliography{refs}{}
\bibliographystyle{aasjournal}

\end{document}